\documentclass[referee,a4paper,12pt,traditabstract]{swsc} 

\usepackage{graphicx}
\usepackage{txfonts}
\usepackage{subfigure}
\usepackage{epstopdf}
\usepackage{lineno}
\usepackage[authoryear,round]{natbib}
\usepackage[backref]{hyperref}
\usepackage{url}

\hypersetup{colorlinks=true,citecolor=cyan,urlcolor=cyan,linkcolor=blue}

\begin{document}

   \title{Coronal hole evolution from multi-viewpoint data as input for a STEREO solar wind speed persistence model}

   
   \titlerunning{STEREO+CH persistence model}

   \authorrunning{Temmer, Hinterreiter, Reiss}

   \author{M. Temmer\inst{1}
          \and
          J. Hinterreiter\inst{1} 
          \and
          M.A. Reiss \inst{1,2}
                   }

   \institute{$^1$ Institute of Physics, University of Graz,
              Universitaetsplatz 5, A-8010 Graz, Austria\\
              $^2$ Space Research Institute, Austrian Academy of Sciences, A-8042 Graz, Austria\\
              \email{\href{mailto:manuela.temmer@uni-graz.at}{manuela.temmer@uni-graz.at}} \\
                     \email{\href{mailto:j.hinterreiter@uni-graz.at}{j.hinterreiter@uni-graz.at}}
                     \email{\href{mailto:reissmar@gmx.at}{reissmar@gmx.at}}
             }

 
  \abstract
   {We present a concept study of a solar wind forecasting method for Earth, based on persistence modeling from STEREO in-situ measurements combined with multi-viewpoint EUV observational data. By comparing the fractional areas of coronal holes (CHs) extracted from EUV data of STEREO and SoHO/SDO, we perform an uncertainty assessment derived from changes in the CHs and apply those changes to the predicted solar wind speed profile at 1~AU. We evaluate the method for the time period 2008--2012, and compare the results to a persistence model based on ACE in-situ measurements and to the STEREO persistence model without implementing the information on CH evolution. Compared to an ACE based persistence model, the performance of the STEREO persistence model which takes into account the evolution of CHs, is able to increase the number of correctly predicted high-speed streams by about 12\%, and to decrease the number of missed streams by about 23\%, and the number of false alarms by about 19\%. However, the added information on CH evolution is not able to deliver more accurate speed values for the forecast than using the STEREO persistence model without CH information which performs better than an ACE based persistence model. Investigating the CH evolution between STEREO and Earth view for varying separation angles over $\sim$25--140$^{\circ}$ East of Earth, we derive some relation between expanding CHs and increasing solar wind speed, but a less clear relation for decaying CHs and decreasing solar wind speed. This fact most likely prevents the method from making more precise forecasts. The obtained results support a future L5 mission and show the importance and valuable contribution using multi-viewpoint data. 
   
    
    
   
   }        

   \keywords{solar wind --
                solar surface --
                interplanetary space
               }

   \maketitle

\section{Introduction}

High-speed solar wind streams are emanated from coronal holes (CHs), where the magnetic field opens into interplanetary space enabling plasma to escape from the Sun. The speed of the enhanced solar wind is positively correlated to the coronal hole area \citep{nolte76}. CHs are rather slowly evolving features in the solar corona (several hours to days), occurring over all latitudes and appearing, due to low plasma density, as dark regions in EUV image data \citep{cranmer09}. We can, therefore, assume that the solar wind speed from the same solar source is persistent in time which was shown with a statistics over 50 years by \cite{owens13}. Based on this, typically single viewpoint in-situ measurements are used to perform persistence models for solar wind speed forecasting, giving very reasonable results \citep[see e.g.][]{owens13}. A similar performance as a persistence model built on ACE (Advanced Composition Explorer) measurements, has the empirical solar wind forecasting (ESWF\footnote{ESWF is a service provided by ESA's Space Situational Awareness Program (SSA) within the Heliospheric Expert Service Center and is operational at the University of Graz since October 2016 (\href{http://swe.uni-graz.at}{swe.uni-graz.at}).}) model which is built on the linear \textit{CH area-solar wind speed} relation using an empirical formula that relates the CH area observed remotely in the EUV wavelength range with the speed at 1~AU \citep[see][]{vrsnak07,rotter12,rotter15}. With this, ESWF is able to give a forecast of solar wind bulk speed at 1~AU distance, having a lead time of about 4~days and an accuracy of $\pm$1~day in the arrival time and $\pm$150~km/s in the impact speed \citep{reiss16}. \cite{jian15} compared, among others, different numerical models running at the Community Coordinated Modeling Center (CCMC), and found uncertainties in the arrival times of 0.5--3 days \citep[see also][]{lee09,gressl14}.

For short time series and during minimum solar activity (2007--2009), very high correlations between solar wind data from the twin STEREO (Solar Terrestrial Relations Observatory) satellites were found for separation angles up to 30$^{\circ}$ \citep{opitz09}. \cite{turner11} concluded that models based on data from spacecraft located behind Earth in their orbit around the Sun (like STEREO-B and since mid of 2015 STEREO-A) perform even better than ACE based persistence models \citep[see also][]{li04}. Hence, by using in-situ measurements from spacecraft located East of Earth, we are able to improve solar wind forecasting at Earth. However, we must not forget that variations in the structure of a CH do have effects on the emanated solar wind. \cite{gomez11} found from using STEREO data, a clear increase in the in-situ measured solar wind speed for a CH that was expanding. Much stronger effects on the persistence of the solar wind flow are caused by coronal mass ejections (CMEs), but are temporally and spatially restricted \citep[see e.g.][]{temmer17}. 

Here, we present a new concept of persistence modeling by adding two important aspects, i) multi-viewpoint observations and ii) taking into account the history and evolution of the solar wind due to changes of its solar sources. With this we increase the lead time of solar wind forecast and improve its performance. We make use of STEREO in-situ measurements, combined with remote sensing EUV observations from two viewpoints (STEREO and Earth view) to assess changes in the areas of CH. Determined changes in the CHs are used to adapt the STEREO based forecast accordingly. To derive the likelihood of persistence of solar wind speed, we first present for the time range 2008--2012 a thorough statistical analysis between STEREO-B persistence results and ACE. We compare the results in a continuous and an event-based approach, with and without CMEs, and relate the derived changes in the CH areas with changes in the measured solar wind speed. Finally, we show the application of the new concept in an operational mode in 2017 using beacon STEREO-A in-situ and EUV, and SDO (Solar Dynamics Observatory) EUV data for real-time solar wind forecasting.

\section{Data and Methods}

\begin{figure}
   \centering
  \includegraphics[width=0.9\columnwidth]{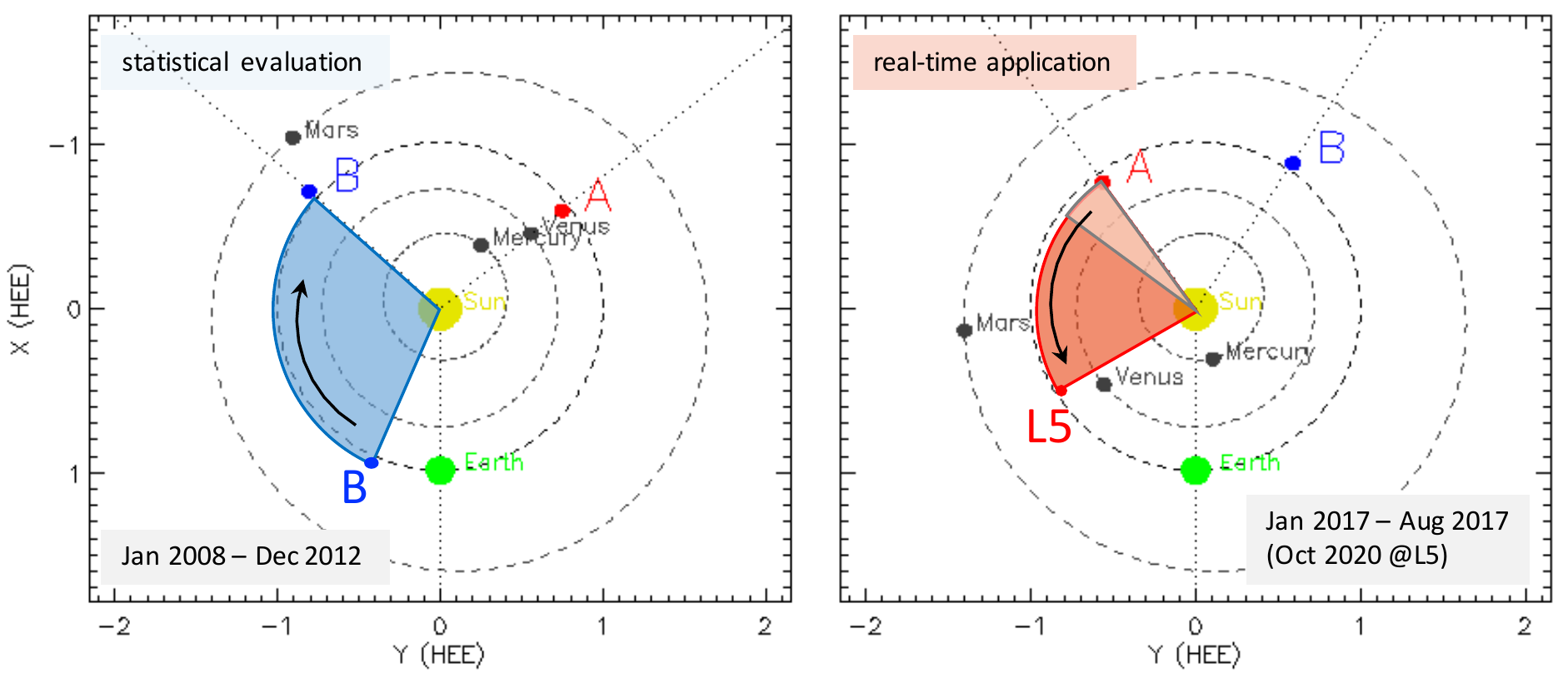}
   \caption{\small Positions of STEREO spacecraft for the time under study. Left: STEREO-B position spans the angular range 23--131$^{\circ}$ East of Earth and is used for the statistical evaluation. Right: Results from the evaluation are used to run the persistence model based on STEREO-A data for the angular range 144--129$^{\circ}$ East and will be monitored further on, with special focus on the L5 position which will be reached end of 2020.} 
   \label{fig:where}
   \end{figure}

Our statistical analysis spans the time range 2008--2012 (to cover with STEREO-B a similar spacecraft position as we currently have with STEREO-A, and to explore a possible L5 mission which corresponds to the time range around autumn 2009 and beginning of 2010) covering the minimum phase of solar cycle 23 and the increasing to maximum phase of solar cycle 24. The statistics will, therefore, comprise time ranges of low and high solar activity that are compared with each other. During 2008--2012, the STEREO-B position covers the angular range 23--131$^{\circ}$ and in 2017, STEREO-A covers 144--129$^{\circ}$, both East of Earth. The year 2008 is very suitable for statistical calibration, due to i) STEREO-B was located in close angular distance to Earth, hence, evolutionary effects of CHs should be small and ii) transient events are sparse due to very low solar activity. We show the application of the new concept to real-time solar wind data from STEREO-A covering the time range January--August in 2017. Figure~\ref{fig:where} illustrates the angles between the spacecraft STEREO-B and STEREO-A with Earth for the time ranges under study (separation angles are given in HEE longitude, neglecting latitude). The real-time forecast of the solar wind speed at Earth has currently (2017) a lead time of about 10 days and will decrease as STEREO-A moves further West (e.g., 4.5 days for the L5 location reached end of 2020).

\subsection{Definition of metrics and persistence models based on in-situ data}\label{data}

We make use of in-situ measurements with 1~min time cadence, from STEREO-B \citep[PLASTIC;][]{galvin08} and ACE \citep[SWEPAM;][]{mccomas98}, to identify streams of enhanced solar wind speed. Before performing the model algorithms and analysis, all in-situ measurements are linearly interpolated onto a 6~h time grid. For the detection of ``events'' we use the same algorithm as \cite{reiss16}: a requirement for a peak to be detected is a minimum speed of 400~km/s before and after the increase, having a peak prominence of at least 60~km/s (which does not need to be monotonically increasing).

A ``hit'' (true positive; TP) is defined if a predicted event overlaps within a time window of $\pm$2~days with a detected event in the reference data set (i.e., ACE). A ``false alarm'' (false positive; FP) is given when an event is predicted but not detected, whereas a ``miss'' (false negative; FN) is a detected event which has no predicted event associated. From the number of hits, false alarms, and misses we calculate well-established skill measures such as the probability of detection (POD), the false alarm ratio (FAR), and the threat score (TS). The TS is a measure of the overall model performance ranging from 0 to 1, where the best TS is 1 and the worst TS is 0. Additionally, the bias (BS) indicates whether the number of observations is underestimated ($BS < 1$) or overestimated ($BS > 1$). This event-based validation approach is supplemented by an error analysis of the predicted solar wind speed timelines for which we compute the arithmetic mean, mean absolute error, and root mean square error, standard deviation and correlation coefficient. For each hit (TP) we compare the predicted and the measured speeds and check whether the solar wind speed was over- or underestimated. In order to make the results comparable, we define the fraction of the underestimated speed (RUS = ratio underestimated speed, with 1$-$RUS = overestimated speed). For a more detailed description on the metrics used we refer the interested reader to the forecast validation tool developed in \cite{reiss16}. As CMEs cause transient peaks in the in-situ measured solar wind speed, we make a separate analysis excluding CMEs from the analysis, for which we use ready catalogs maintained by Richardson \& Cane for ACE \citep[194 CMEs for 2008--2012; see][for a description of the catalog]{richardson10} and for STEREO-B from L.~Jian \citep[189 CMEs for 2008--2012; see][for a description of the catalog]{jian06}. We note that for the real-time application in 2017, STEREO-A data are used and that CMEs were not excluded. 


The realization of a forecast from a persistence model depends on the values measured at earlier time steps given by the separation angle between measured and forecast position (solar angular rotation speed is about 13.2$^\circ$/day). We present three different persistence models that are evaluated against actual ACE measurements: \textit{ACE+27}, based on ACE data forward shifted in time by 27.2753~days \citep[cf.][]{owens13}; \textit{STEREO} persistence, based on STEREO data dynamically forward shifted, according to the angle between the STEREO-B (STEREO-A) spacecraft and Earth; the new concept model \textit{STEREO+CH}, based on the STEREO persistence model taking into account the evolution of CH.

\subsection{EUV data reduction and coronal hole evolution}

Variations in the sources of enhanced solar wind speed, i.e., CHs, cause changes in the measured bulk velocity at 1~AU. To calculate these variations, we extract fractional CH areas from EUV data in STEREO \citep[EUVI;][]{wuelser04}, SoHO \citep[EIT;][]{delaboudiniere95}, and SDO \citep[AIA;][]{lemen12}. The fractional areas, $A(t)$, are derived within a meridional slice of $\pm 7.5^\circ$ around the central meridian, for which a linear relation with the measured solar wind speed after 4~days at 1~AU exists \cite[for more details see][]{vrsnak07,rotter12}. For the extraction we utilize 1024$\times$1024 image data provided by the SoHO/EIT instrument in the 195\AA~band (2008-2010) the SDO/AIA instrument in the 193\AA~band (2010-2012, 2017) with a time cadence of 6~hours, and the STEREO/SECCHI/EUVI instrument in the 195\AA~band (STEREO-B: 2008--2012, STEREO-A: 2017 - beacon/quick-look data) with a time cadence of 1~hour. To match the in-situ data, all EUV data are linearly interpolated onto a 6~h time grid.

Using the same extraction technique as for ESWF (a histogram-based segmentation method from the intensity distribution in the full-disk EUV images), we apply a threshold value of $f \times$median on-disk intensity, where $f=0.35$ for AIA, $f=0.47$ for EIT (AIA and EIT areas were cross-checked for the overlapping time range in 2010), and $f=0.32$ for STEREO. For CHs spanning their areas beyond $\pm 60^\circ$ heliographic latitude and longitude, we use an additional multiplication factor of 1.6 (to take into account projection effects as CHs appear less dark close to the limb). For further details see \cite{reiss16}.  


Based on this we define the ``CH evolution ratio'' r$_{\rm CH}$ by, 
 \begin{equation}
      r_{\rm CH} = A_{\rm ST} / (A_{\rm Earth}+A_{\rm ST}) \,,
   \end{equation}
where A$_{\rm ST}$ (A$_{\rm Earth}$) refers to the fractional CH areas as derived from STEREO (Earth view, SDO or SoHO). To get a more smooth variation in the CH evolution ratio we use a cutoff for A$_{\rm ST, Earth}<$0.02. From this we compute the median $r_{\rm CH_{med}}$. Deriving $r_{\rm CH} \sim r_{\rm CH_{med}}$ we can assume that the CH fractional areas stay rather constant, hence, reveal a low degree of evolution. For $r_{\rm CH} \lessgtr r_{\rm CH_{med}}$ the solar wind sources underwent some changes and due to expanding/decaying CH areas the persistence model may under-/over-estimate the solar wind speed in its forecast for Earth. We note that $r_{\rm CH}$ cannot be calculated for 21\% of the time range due to gaps in EUV image data (data gaps from EIT: 15\%; AIA: 8\%; 2\% overlapping time). All derived values of $r_{\rm CH}$ are shifted in time by $+$4~days in order to match the corresponding in-situ measurements.

\section{Investigating coronal hole evolution and response in solar wind speed}\label{CH:dist}

\begin{figure}[h]
   \centering
  \includegraphics[width=1\columnwidth]{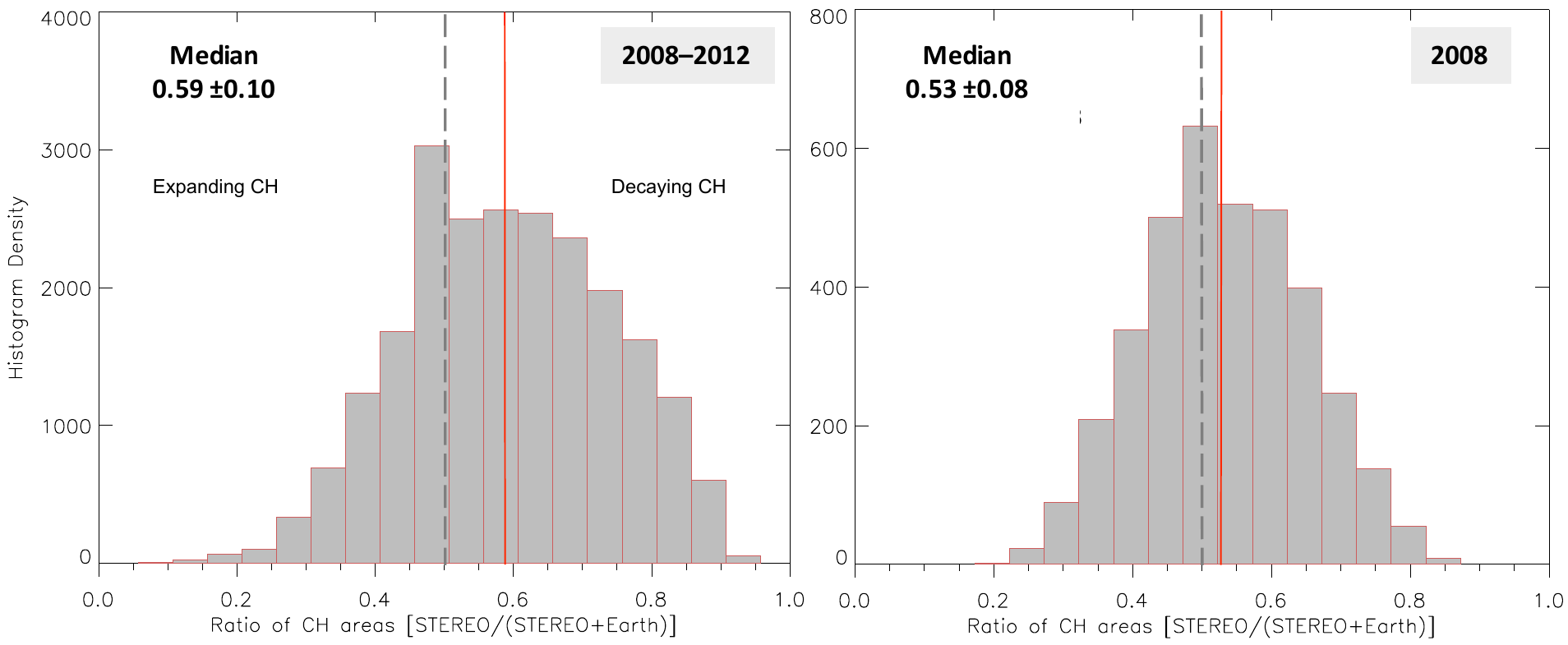}
   \caption{\small Histogram for the CH evolution ratio covering 2008--2012 (left panel) and separately for the year 2008 (right panel). The derived median is given by a red vertical line. } 
   \label{fig:histo}
   \end{figure}

Figure~\ref{fig:histo} shows the distribution of $r_{\rm CH}$ over the entire time range under study and separately for 2008. For the calibration year 2008, we derive a rather symmetric distribution with a median of 0.53, giving evidence that most of CH areas evolve slowly over time spans of $\sim$2 days and during low solar activity phases (as expected). The distribution of $r_{\rm CH}$ over the entire time range is rather asymmetric with a median of 0.59, referring to an evolutionary effect in CH areas. As the spacecraft separation (and solar activity) increases, we find a higher rate of decaying than expanding fractional CH areas (cf.\,results in Section~\ref{CH:speed}). Either the decay phase of CHs is of longer duration compared to the expansion phase, or as the solar activity picked up more CHs decayed. Clearly, persistence models based on future L5 missions, need to be updated/corrected as the solar wind sources may evolve over time spans of $\sim$4.5 days.

To assess different levels of persistence (i.e.\,how strongly CHs evolve), we calculate for each year separately the median $r_{\rm CH_{med}}$ and its median absolute deviation (MAD) $r_{\rm CH_{MAD}}$. In order to relate the derived distribution of $r_{\rm CH}$ to differences in the measured in-situ solar wind speed between ACE and STEREO, we calculate the differences between the measured solar wind bulk speed $v=v_{\rm ACE}-v_{\rm STEREO}$, the median $v_{\rm med}$ and its $v_{\rm MAD}$. All values are given separately for each year in Table~\ref{tab:CHrat}, together with the arithmetic mean and standard deviation as well as the standard error of the mean and median. On average, the differences in the solar wind speed between the two spacecraft show no clear trend of increased values for larger separation angles. Compared to that, the CH evolution ratio increases constantly from 2008 to 2012 (median from 0.53 to 0.63; cf.\,results in Section~\ref{CH:dist}).

   \begin{table}
      \caption[]{Statistical results (median, MAD, standard error of the median ($\mathit{SE}_{\rm med}$), arithmetic mean, standard error of the mean ($\mathit{SE}_{\rm mean}$), standard deviation) of solar wind speed differences, $v$ in km/s derived from ACE$-$STEREO (CMEs excluded), and CH evolution ratio values, $r_{\rm CH}$, separately listed for each year covering 2008--2012 as well 2017. }
         \label{tab:CHrat}
     $$ 
         \begin{array}{l|cccc|cccc}
            \hline
            \noalign{\smallskip}
            Year &  v_{\rm med}  \pm v_{\rm MAD} & {\mathit{SE}_{\rm med}} & \tilde{v} \pm 1\sigma &  {\mathit{SE}_{\rm mean}} & r_{\rm CH_{med}} \pm r_{\rm CH_{MAD}} & {\rm \mathit{SE}_{med}} & \tilde{r_{\rm CH}} \pm 1\sigma & {\rm \mathit{SE}_{mean}} \\
            \noalign{\smallskip}
            \hline
            \noalign{\smallskip}
            2008   	&  1.52  \pm 37.57	& 2.18	&  8.13 \pm 66.30   &  1.74 & 0.534 \pm 0.083 & 0.004 & 0.536 \pm 0.114 & 0.003		\\
            2009	&  12.08 \pm 36.49	& 1.99	&  13.16 \pm 60.88 	&  1.59 & 0.559 \pm 0.088 & 0.004 & 0.563 \pm 0.122 & 0.003		\\
            2010	&  7.68 \pm  49.46	& 2.93	&  8.78 \pm 89.24	&  2.34	& 0.605 \pm 0.114 & 0.005 & 0.601 \pm 0.156 & 0.004		\\
            2011	&  5.70 \pm  48.80	& 3.17	&  7.00 \pm 96.68 	&  2.53 & 0.609 \pm 0.109 & 0.005 & 0.604 \pm 0.141 & 0.004		\\
            2012	&  9.31 \pm  51.12	& 2.98  &  11.51 \pm 90.82	&  2.38 & 0.632 \pm 0.124 & 0.005 & 0.625 \pm 0.158 & 0.004		\\
            2017	&  7.74 \pm  63.01	& 4.30	&  11.80 \pm 113.87	&  3.43 & 0.508 \pm 0.101 & 0.005 & 0.524 \pm 0.145 & 0.005		\\
            \noalign{\smallskip}
            \hline
         \end{array}
     $$ 
   \end{table}


\subsection{Relating CH evolution ratio to solar wind speed}\label{CH:speed}
  
\begin{figure}
   \centering
  \includegraphics[width=1.\columnwidth]{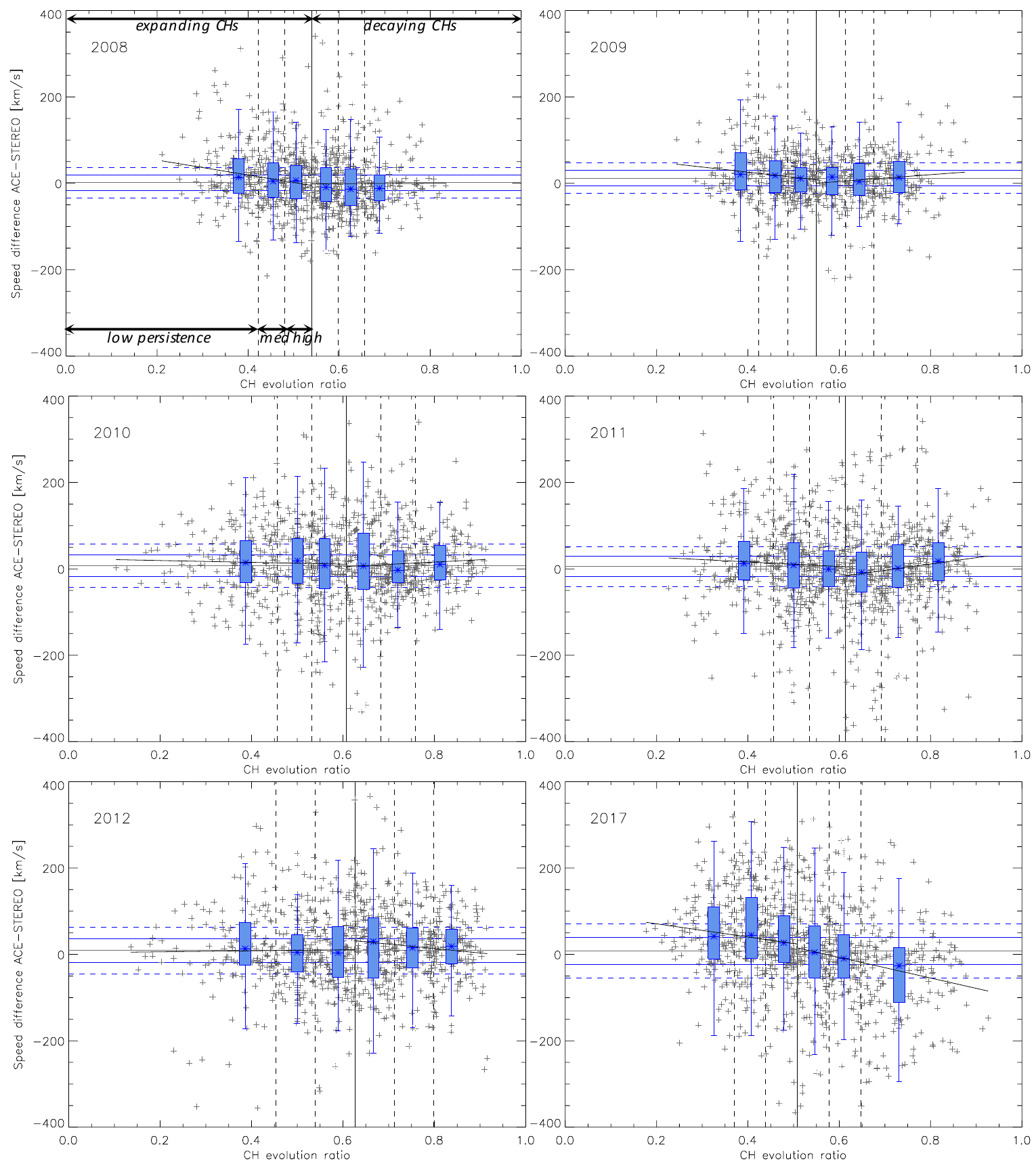}
   \caption{\small CH evolution ratio versus solar wind speed differences between ACE and STEREO in-situ measurements (CME events excluded). To show the overall trend, we overplot a linear fit (black line) to the data of the category expanding CHs and separately for decaying CHs. We mark the median of the CH evolution ratio of each year (vertical solid line) and give for each performance class the calculated MAD ranges (vertical dashed lines), for high performance $\pm$0.7MAD, for medium performance $\pm$1.4MAD, and for low performance outside $\pm$1.4MAD. Within each category we give a box-and-whisker plot showing the median (blue asterisk), the difference between the lower and upper quartile (blue box; interquartile range or IQR), and the upper and lower whisker given by 1.5$\times$IQR. Horizontal lines refer to the speed median (gray) and MAD ranges with $\pm$0.5MAD (blue solid), $\pm$1MAD (blue dashed). Note that CMEs are not excluded for 2017.} 
   \label{fig:ratio}
   \end{figure}

Figure~\ref{fig:ratio} shows for the years 2008--2012 and 2017, the CH evolution ratio for decaying ($r_{\rm CH}>r_{\rm CH_{med}}$) and expanding CHs ($r_{\rm CH}<r_{\rm CH_{med}}$) versus $v$. A linear fit, separately calculated for expanding CHs and decaying CHs, shows that, on average expanding CHs tend to be related to a positive speed difference, i.e.\,an increase in the solar wind speed (as would be expected). However, we note that this trend is not very distinct. The years 2008 and 2017 reveal the most obvious relation between expanding CH areas and positive speed difference. In comparison, decaying CHs seem to be related to negative as well positive speed differences. There is no consistent trend found for decaying CHs to be related to solar wind streams of decreasing speed. We also note that results from 2009 show the least scattering maybe due to the deep solar minimum and the flat heliospheric current sheet.

This (weak) trend enables to give for expanding CHs an uncertainty estimate in the forecast speed of the persistence model, that should improve hit/miss rates. We define different persistence levels for $r_{\rm CH}$, based on the derived MAD values (as robust measure and less sensitive to outliers). \textit{high persistence}: $r_{\rm CH_{med}}-0.7 r_{\rm CH_{MAD}} \leq r_{\rm CH} < r_{\rm CH_{MAD}}$; \textit{medium persistence}: $r_{\rm CH_{med}}-1.4 r_{\rm CH_{MAD}} \leq r_{\rm CH} < r_{\rm CH_{med}}-0.7 r_{\rm CH_{MAD}}$; \textit{low persistence}: $r_{\rm CH_{med}}-1.4 r_{\rm CH_{MAD}} < r_{\rm CH}$. As shown in Figure~\ref{fig:ratio}, within each level we visually display the derived median values for the speed and the speed distribution as a box-and-whisker plot (for completion also for decaying CHs). On average, the interquartile range of the three different persistence levels for expanding CHs spans speed differences up to about $+$1.5$v_{\rm MAD}$. Each persistence level, high/medium/low, is then related to speed uncertainties of $+$0.5$v_{\rm MAD}$/$+$1.0$v_{\rm MAD}$/$+$1.5$v_{\rm MAD}$. With this we assess the speed uncertainty that we expect in the solar wind speed forecast when the CH fractional area undergo changes.

\subsection{STEREO+CH persistence model}

\begin{figure}
   \centering
  \includegraphics[width=1.\columnwidth]{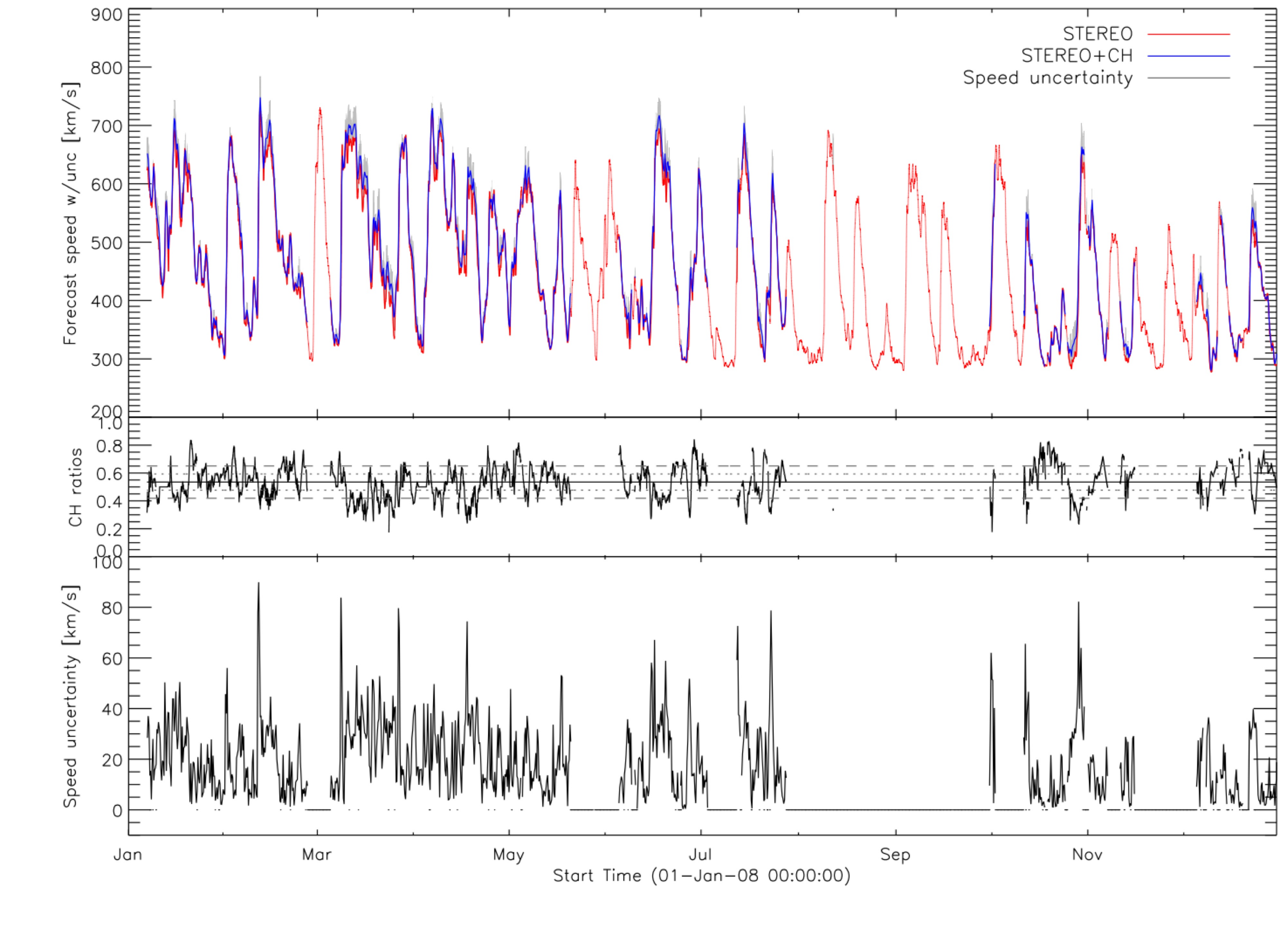}
   \caption{\small For the year 2008 we show in the top panel the results from the STEREO persistence model (red profile), the derived speed uncertainties using CH evolution information (gray profile), and the new STEREO+CH forecast (blue profile) - for time ranges with EUV data gaps no speed uncertainty can be given and the STEREO profile is used as forecast (thin red profile); in the middle panel we give the evolution of the CH ratio as derived from EUV data together with its median and $+$0.7/$+$1.4MAD levels (vertical solid, dotted and dashed lines); in the bottom panel the speed uncertainties are shown.}
   \label{fig:test}
   \end{figure}

\begin{figure}
   \centering
  \includegraphics[width=1.\columnwidth]{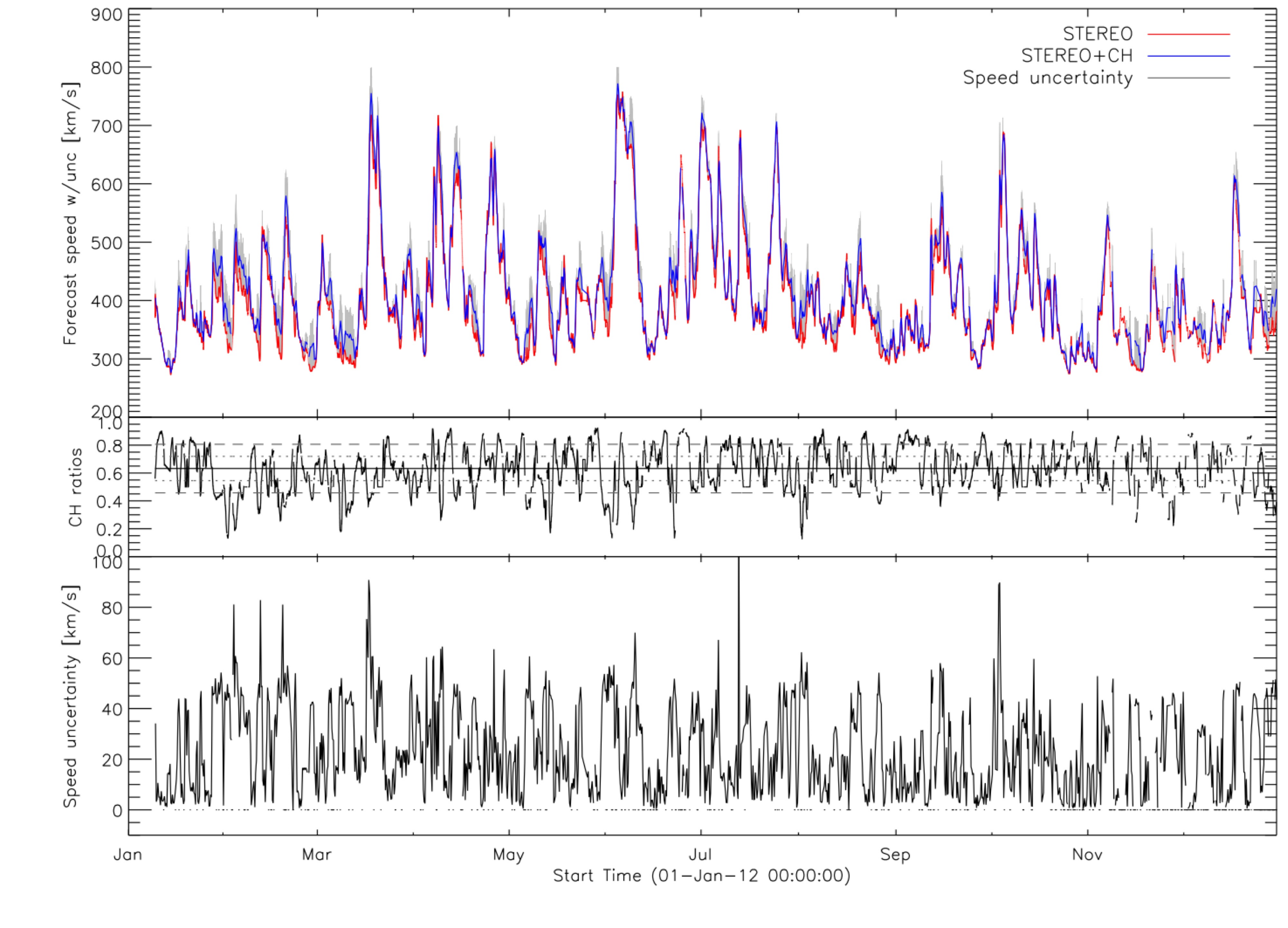}
   \caption{\small Same as Figure~\ref{fig:test} but for the year 2012.} 
   \label{fig:test2012}
   \end{figure}

To get a solar wind speed forecast that is updated with the information about a possible evolution of the source of the solar wind stream, we use the STEREO persistence model and modify its results. According to the obtained trend between expanding CHs and positive speed difference (cf.\,Figure~\ref{fig:ratio}), we simply increase the derived solar wind forecast from STEREO during those time ranges which are related to expanding CHs ($r_{\rm CH}<r_{\rm CH_{med}}$). As described in the previous section, for high persistence levels we increase the derived STEREO forecast speed by $+0.5 v_{\rm MAD}$, for medium persistence we increase by $+1 v_{\rm MAD}$, for low persistence we increase the speed by $+1.5 v_{\rm MAD}$ (cf.\,Table~\ref{tab:CHrat}). Considering that decaying CHs affect the solar wind speed in a less clear trend than expanding CHs, no change is applied for time ranges revealing $r_{\rm CH}>r_{\rm CH_{med}}$. As these uncertainties represent an upper limit of the forecast speed, we calculate the arithmetic mean between the maximum envelope of the modified speed profile and the original STEREO persistence forecast. This completes the STEREO+CH persistence model from which we derive a new forecast of solar wind speed having intensified and slightly broader speed profiles for those streams that are related to expanding CHs. Data gaps due to the lack of EUV data are filled by the original results from the STEREO persistence model. Figures~\ref{fig:test} and~\ref{fig:test2012} show for 2008 and 2012 the STEREO solar wind speed forecasting curve (red line) together with the estimated speed uncertainties (gray line) from the CH evolution and the resulting forecast curve, the STEREO+CH persistence model (blue line). The evolution of the CH ratio area and estimated speed uncertainties are given in the middle and bottom panels, respectively. We note that the STEREO persistence model has a variable lead time depending on the spacecraft location - as studied here, for up to 10.9~days. In comparison, the derived uncertainties for the STEREO+CH persistence model are based on the information of CH evolution (EUV data from Earth view are compared to STEREO), and therefore restricted to a lead time of 4~days.  

\section{Performance comparison between different models}

\begin{figure}
   \centering
  \includegraphics[angle=90,width=0.9\columnwidth]{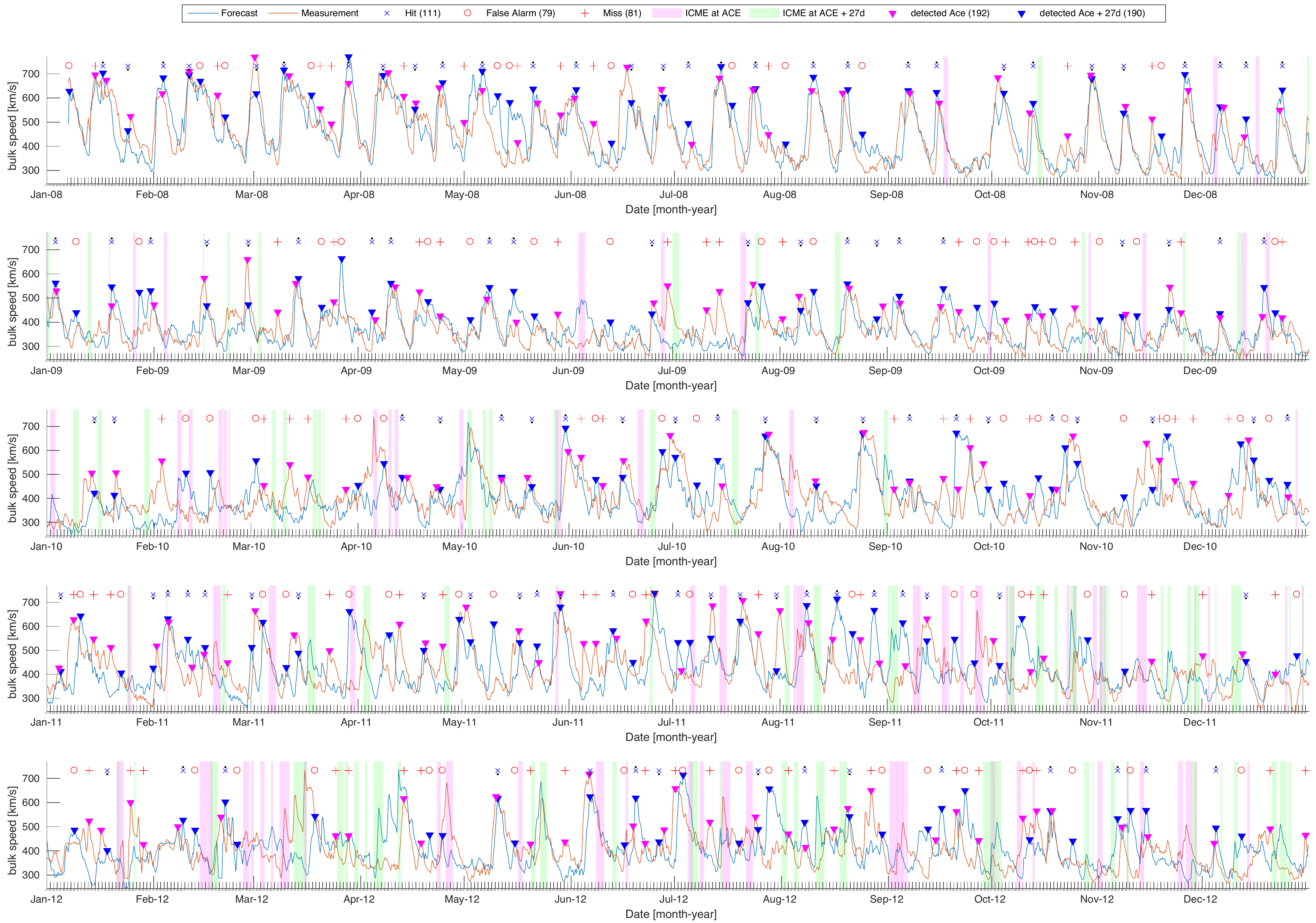}
   \caption{\small Solar wind speed measured by ACE (red curve) covering the time range 2008--2012 with the forecast from the ACE+27 persistence model overlaid (blue curve). Extracting the R\&C list, we mark detected CME events at ACE (magenta; ICME at ACE) and during the time span covered by the persistence model ACE+27, i.e.\,during the subsequent rotation (green; ICME at ACE+27). Triangles denote the detected peaks in each profile, which are labeled as a hit (blue cross), false alarm (red circle), or miss (red cross). Black dots below or above a hit symbol marks that the actually measured speed was under- or overestimated by the forecast. } 
   \label{fig:p27}
   \end{figure}

\begin{figure}
   \centering
  \includegraphics[angle=90,width=.9\columnwidth]{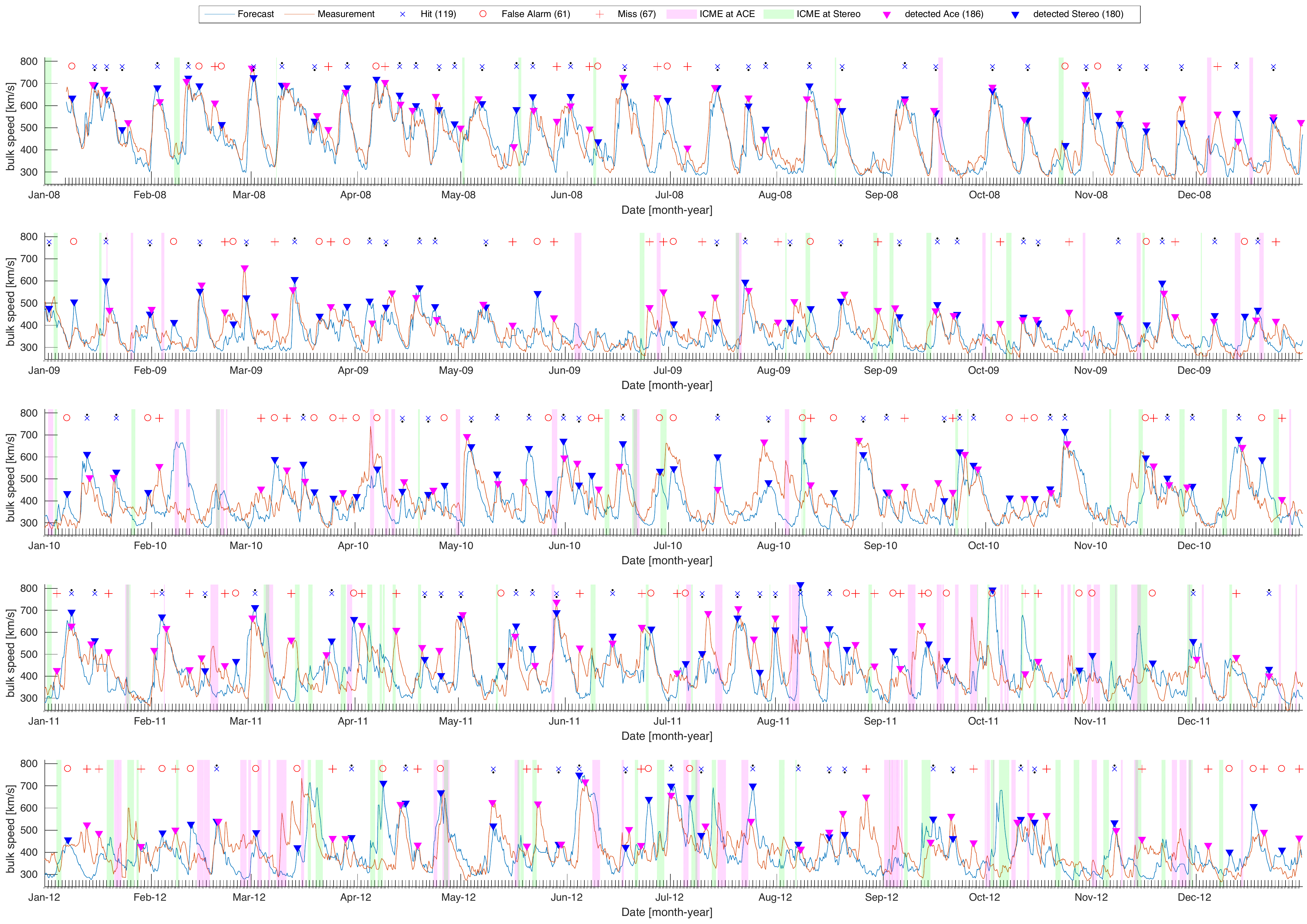}
   \caption{\small Same as Figure~\ref{fig:p27}, but overlaid with STEREO-B persistence model forecast results. Detected CME events at ACE (R\&C list) and STEREO-B (Jian list) are marked in magenta and green. } 
   \label{fig:p4}
   \end{figure}

\begin{figure}
   \centering
  \includegraphics[angle=90,width=.9\columnwidth]{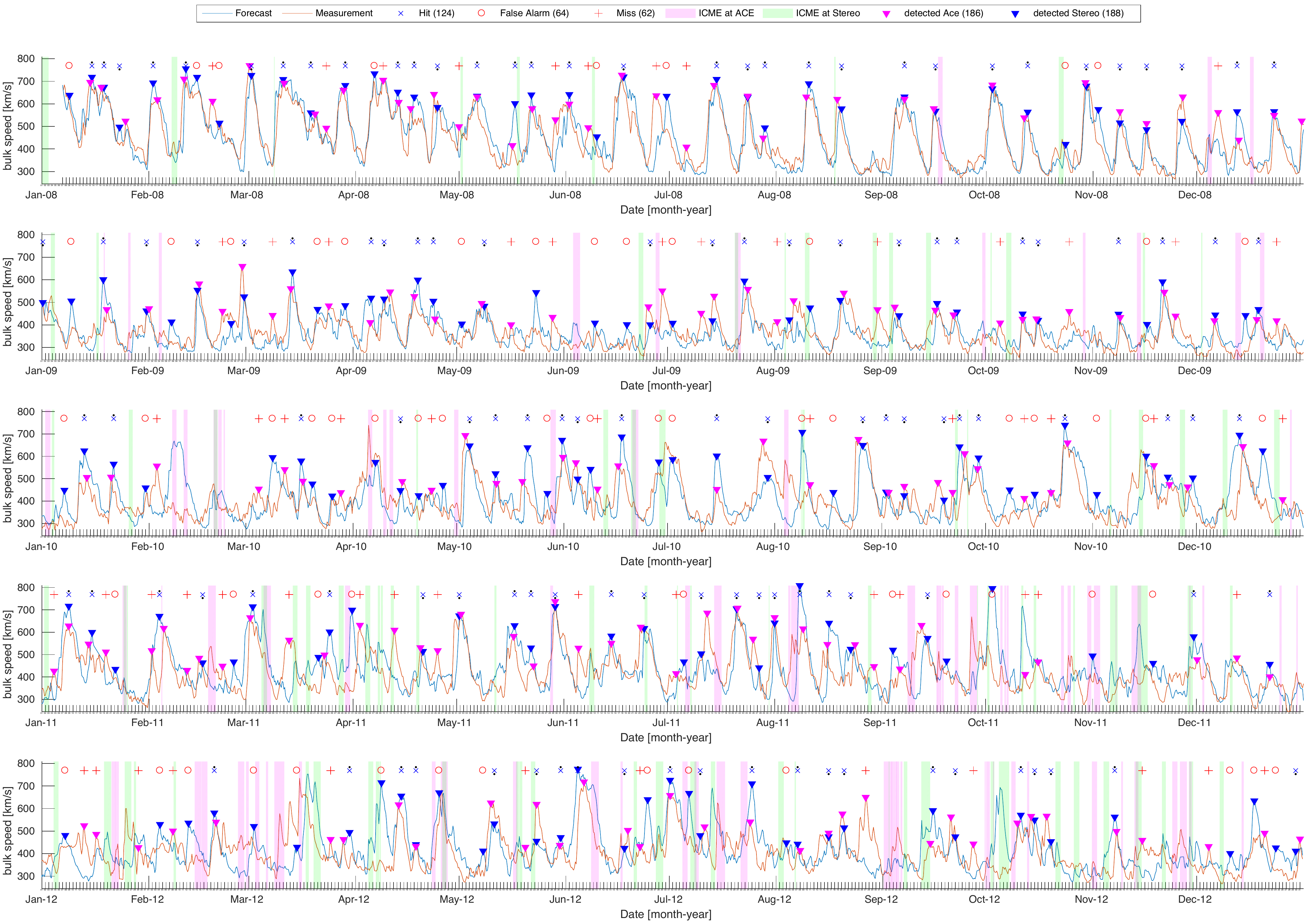}
   \caption{\small Same as Figure~\ref{fig:p4}, but overlaid with STEREO+CH model results. } 
   \label{fig:ST-CH}
   \end{figure}

\begin{figure}
   \centering
  \includegraphics[width=1.\columnwidth]{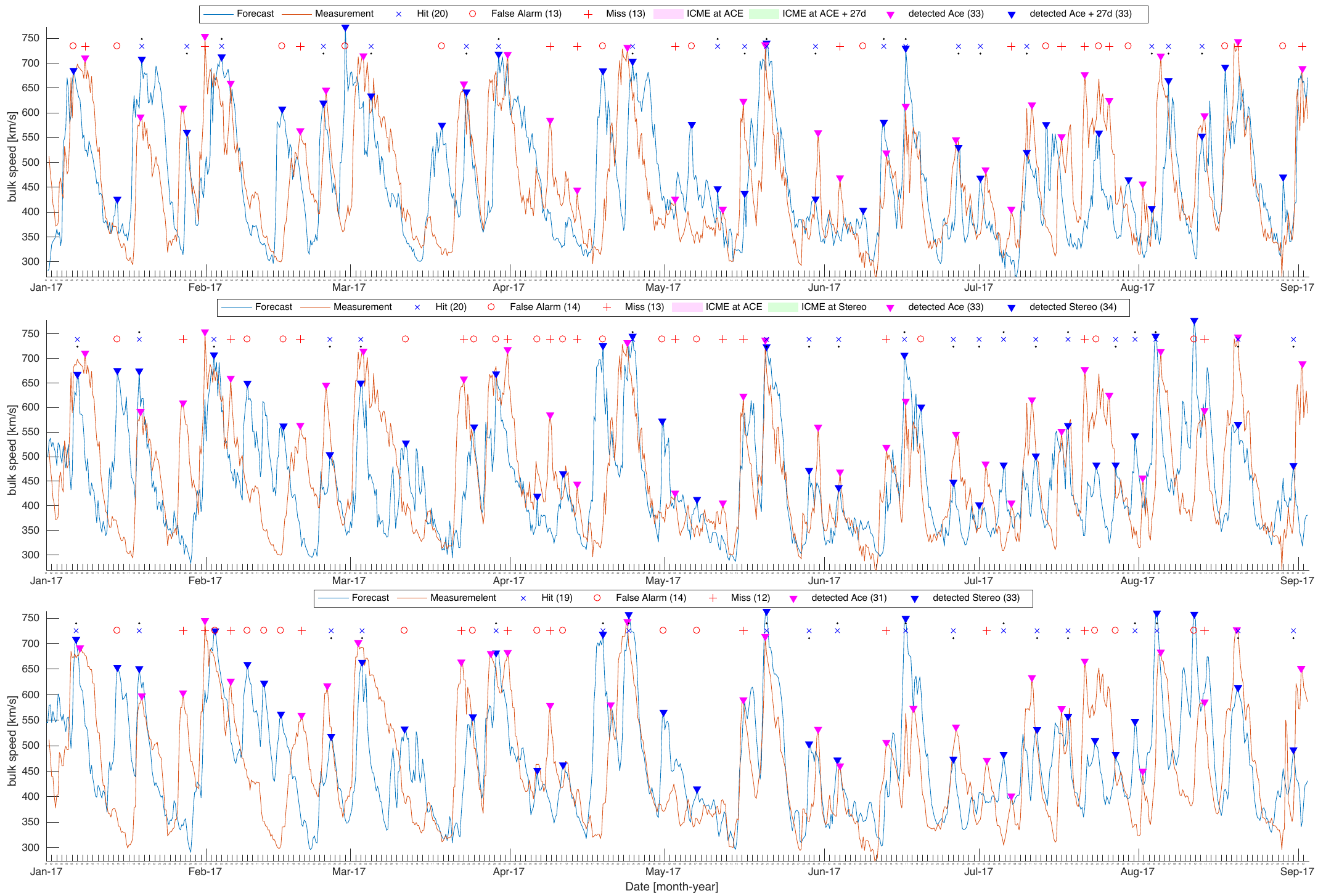}
   \caption{\small Results for the year 2017 using ACE+27 (top), STEREO (middle), and STEREO+CH (bottom) persistence with the same symbols and description as given for Figures~\ref{fig:p4} and~\ref{fig:ST-CH}. } 
   \label{fig:2017}
   \end{figure}

We present the performance of ACE+27, STEREO, and STEREO+CH persistence in comparison to ACE measurements using an event-based and continuous forecast validation approach (see Section~\ref{data}). Figures~\ref{fig:p27}, \ref{fig:p4}, and \ref{fig:ST-CH} show for the time range 2008--2012 the evolution of the solar wind bulk speed measured at ACE in comparison to the forecasts from the ACE+27, STEREO, and STEREO+CH persistence model, respectively. The number of identified peaks in the speed profiles, together with the number of hits, false alarms, and misses are given in the legend of each plot. Time ranges during which CMEs occurred are marked. With black dots that are put below or above a hit (blue crosses), we show that the peak speed was under- or overestimated by the forecast. For comparison, we present in Figure~\ref{fig:2017} the real-time application (using beacon data) for the three models during 2017 (January-August)\footnote{These results use the $r_{\rm CH_{med}}$ as derived over the year 2017. For the real-time application, this $r_{\rm CH_{med}}$ is adapted every month.}. The results for the time range under study, and separately for the years 2008, 2012, and 2017 are summarized in Tables~\ref{tab:events} and~\ref{tab:cont}. 

Inspecting Table~\ref{tab:events} we obtain, compared to ACE+27, in general, a better performance of the models STEREO and STEREO+CH with the highest values for POD and lowest for FAR. For comparison, we give the results also for CMEs included in the statistics. For 2008 almost no changes are revealed in the performance of the models by excluding the few (3) CME events which occurred. Interestingly, for 2012 a slightly better performance is revealed when CMEs are included, that might be related to CMEs occurring close to CHs \cite[see e.g.,][]{dhuys14}. As we focus in our study on solar wind streams, we describe the performance of the models derived by excluding CMEs from the time series. In total (2008--2012) the number of hits can be increased from 111 (ACE+27) to 119 (STEREO) and 124 (STEREO+CH), while the number of misses can be decreased from 81 (ACE+27) to 67 (STEREO) and 62 (STEREO+CH). Also, the false alarms are fewer for STEREO (61) and STEREO+CH (64) compared to ACE+27 (79). Inspecting TS values, we obtain an improvement from ACE+27 (TS=0.41) to STEREO (TS=0.49) and STEREO+CH (TS=0.50). ACE+27 looses 51\% performance comparing low and high solar activity (TS=0.57 in 2008 and 0.28 in 2012), while for STEREO it is 42\% (0.67/0.39 in 2008/2012) and STEREO+CH 30\% (0.65/0.46 in 2008/2012). Results for the continuous values of the measured and predicted solar wind speed, as their mean and median values, errors, standard deviations and correlation coefficients, are given in Table~\ref{tab:cont} and show that STEREO and STEREO+CH are able to forecast the measured speed profile better than ACE+27, with RMSE of the order of 80~km/s for STEREO and STEREO+CH, and 88~km/s for ACE+27. On average, for ACE+27 around 46\% of the detected hits underestimate the peak speed, 45\% for STEREO and 42\% for STEREO+CH. Using STEREO and STEREO+CH we obtain a correlation coefficient of 0.72 and 0.71, respectively, between measured and predicted solar wind speed while lower, 0.56, for ACE+27. When investigating the entire time span (2008--2012), by excluding CMEs from the time series the correlation coefficient increases for all models. The accuracy of the arrival time for each hit lies for all models in the range of $\pm$1~day.

\begin{table}
      \caption[]{Event based statistics for ACE+27, STEREO and STEREO+CH persistence models covering the time range 2008--2012 and 2017 (January--August). ALL = 2008--2012; YES = CMEs occurrences are included in the statistical analysis; NO = CMEs are excluded from the statistical analysis; Total = number of all predicted peaks; TP = hit; FP = false alarm; FN = miss; FAR = False Alarm Ratio; TS = Threat Score (worst: TS = 0, best: TS = 1); BS = Bias (BS $<$ 1 underestimation; BS $>$ 1 overestimation); POD = Probability of Detection; 

}
         \label{tab:events}
     $$ 
         \begin{array}{lllllllll}
\hline
	 Data & Total & \mathit{TP} & \mathit{FP} & \mathit{FN} & \mathit{FAR} & \mathit{TS} & \mathit{BS} & \mathit{POD} \\ \hline
	\multicolumn{9}{c}{\mathit{ACE}+27}  \\ \hline
	2008/YES & 41 & 30 & 11 & 12 & 0.29 & 0.57 & 0.98 & 0.73 \\ \hline
	2008/NO & 41 & 30 & 11 & 12 & 0.29 & 0.57 & 0.98 & 0.73 \\ \hline
	2012/YES & 44 & 22 & 22 & 22 & 0.50 & 0.33 & 1.00 & 0.50 \\ \hline
	2012/NO & 33 & 15 & 18 & 21 & 0.58 & 0.28 & 0.92 & 0.45 \\ \hline
	ALL/YES & 215 & 126 & 89 & 90 & 0.42 & 0.41 & 1.00 & 0.59 \\ \hline
	ALL/NO & 190 & 111 & 79 & 81 & 0.42 & 0.41 & 0.99 & 0.58 \\ \hline
	2017/YES & 30 & 18 & 12 & 13 & 0.42 & 0.42 & 0.97 & 0.60 \\ \hline
   	\multicolumn{9}{c}{\mathit{STEREO}~persistence~model}  \\ \hline
	2008/YES & 42 & 34 & 8 & 8 & 0.19 & 0.68 & 1.00 & 0.81 \\ \hline
	2008/NO & 41 & 33 & 8 & 8 & 0.20 & 0.67 & 1.00 & 0.80 \\ \hline
	2012/YES & 43 & 25 & 18 & 19 & 0.43 & 0.40 & 0.98 & 0.58 \\ \hline
	2012/NO & 30 & 18 & 12 & 16 & 0.47 & 0.39 & 0.88 & 0.60 \\ \hline
	ALL/YES & 208 & 137 & 71 & 79 & 0.37 & 0.48 & 0.96 & 0.66 \\ \hline
	ALL/NO & 180 & 119 & 61 & 67 & 0.36 & 0.49 & 0.97 & 0.66 \\ \hline
	2017/YES & 33 & 20 & 13 & 11 & 0.35 & 0.45 & 1.06 & 0.61 \\ \hline
	\multicolumn{9}{c}{\mathit{STEREO+CH}~persistence~model}  \\ \hline
	2008/YES & 41 & 33 & 8 & 9 & 0.21 & 0.66 & 0.98 & 0.80 \\ \hline  
	2008/NO & 40 & 32 & 8 & 9 & 0.22 & 0.65 & 0.98 & 0.80 \\ \hline
	2012/YES & 47 & 29 & 18 & 15 & 0.34 & 0.47 & 1.07 & 0.62 \\ \hline 
	2012/NO & 36 & 22 & 14 & 12 & 0.35 & 0.46 & 1.06 & 0.61 \\ \hline
	ALL/YES & 214 & 143 & 71 & 73 & 0.34 & 0.50 & 1.00 & 0.67 \\ \hline
	ALL/NO & 188 & 124 & 64 & 62 & 0.33 & 0.50 & 1.01 & 0.66 \\ \hline
	2017/YES & 33 & 19 & 14 & 12 & 0.39 & 0.42 & 1.06 & 0.58 \\ \hline
\end{array}
     $$ 
   \end{table}

   \begin{table}
      \caption[]{Statistics on continuous values for solar wind speed [km/s] covering the time range 2008--2012 and 2017 (January--August). ALL = 2008--2012; YES = CMEs occurrences are included in the statistical analysis; NO = CMEs are excluded from the statistical analysis; ME = mean error; MAE = mean average error; RMSE = root mean square error; meanP = mean speed predicted; stddevP = standard deviation predicted; meanM = mean speed measured; stddevM = standard deviation measured; CC = correlation coefficient; RUS = ratio of underestimated speed (calculated for hits only);

}
         \label{tab:cont}
     $$ 
         \begin{array}{lrlllrrrlr}
\hline
	Data  & \mathit{ME} & \mathit{MAE}  & \mathit{RMSE}  & meanP  & stddevP  & meanM  & stddevM  & \mathit{CC}  & \mathit{RUS}   \\ \hline
	\multicolumn{9}{c}{ACE+27}  \\ \hline
	2008/YES & -6.62 & 57.96 & 77.13 & 458.96 & 116.48 & 452.34 & 114.89 & 0.67 & 0.33 \\ \hline
	2008/NO & -6.95 & 58.07 & 77.31 & 460.10 & 116.75 & 453.15 & 115.44 & 0.67 & 0.33 \\ \hline
	2012/YES & -0.29 & 73.44 & 98.92 & 407.90 & 81.91 & 407.61 & 82.38 & 0.31 & 0.45 \\ \hline
	2012/NO & -0.68 & 71.13 & 95.39 & 408.81 & 82.78 & 408.13 & 81.49 & 0.49 & 0.40 \\ \hline
	ALL/YES & -1.84 & 67.32 & 90.23 & 412.92 & 94.53 & 411.08 & 93.67 & 0.53 & 0.45 \\ \hline
	ALL/NO & -2.36 & 66.03 & 88.31 & 414.33 & 96.22 & 411.97 & 94.83 & 0.56 & 0.46 \\ \hline
	2017/YES & -1.43 & 81.86 & 108.41 & 462.42 & 113.20 & 460.99 & 110.95 & 0.62 & 0.61 \\ \hline
    \multicolumn{9}{c}{STEREO~persistence~model}  \\ \hline
	2008/YES & 8.26 & 46.39 & 63.79 & 444.08 & 116.20 & 452.34 & 114.89 & 0.80 & 0.59 \\ \hline
	2008/NO & 8.13 & 46.16 & 63.73 & 445.62 & 117.02 & 453.75 & 115.96 & 0.78 & 0.58 \\ \hline
	2012/YES & 6.54 & 69.90 & 94.44 & 401.06 & 93.64 & 407.61 & 82.38 & 0.54 & 0.44 \\ \hline
	2012/NO & 11.62 & 66.00 & 88.27 & 396.62 & 91.37 & 408.24 & 82.64 & 0.71 & 0.44 \\ \hline
	ALL/YES & 8.69 & 59.46 & 82.71 & 402.39 & 99.02 & 411.08 & 93.67 & 0.61 & 0.47 \\ \hline
	ALL/NO & 9.79 & 57.15 & 79.28 & 402.15 & 99.71 & 411.94 & 95.08 & 0.72 & 0.45 \\ \hline
	2017/YES & 14.22 & 87.26 & 116.60 & 446.77 & 101.59 & 460.99 & 110.95 & 0.61 & 0.65 \\ \hline
	\multicolumn{9}{c}{STEREO+CH~persistence~model}  \\ \hline
	2008/YES & -1.70 & 46.78 & 63.16 & 454.05 & 121.25 & 452.34 & 114.89 & 0.80 & 0.39 \\ \hline
	2008/NO & -1.94 & 46.53 & 63.01 & 455.69 & 122.08 & 453.75 & 115.96 & 0.77 & 0.38 \\ \hline
	2012/YES & -12.17 & 72.84 & 96.85 & 419.77 & 97.20 & 407.61 & 82.38 & 0.53 & 0.45 \\ \hline
    2012/NO & -7.00 & 68.58 & 90.24 & 415.24 & 95.30 & 408.24 & 82.64 & 0.62 & 0.45 \\ \hline
	ALL/YES & -5.12 & 60.45 & 83.25 & 416.20 & 102.31 & 411.08 & 93.67 & 0.62 & 0.44 \\ \hline
    ALL/NO & -3.86 & 57.99 & 79.62 & 415.80 & 103.05 & 411.94 & 95.08 & 0.71 & 0.42 \\ \hline
	2017/YES & -6.60 & 84.88 & 113.60 & 467.59 & 104.07 & 460.99 & 110.95 & 0.64 & 0.42 \\ \hline
\end{array}
     $$ 
   \end{table}

\section{Summary and discussion}

We present the concept of a new persistence model (STEREO+CH) to forecast solar wind speed using the advantage of multi-viewpoint satellite data. The model is based on a STEREO persistence method using STEREO-B (and -A in 2017) in-situ measurements shifted forward by a variable time span ($\sim$2--10 days) according to the angle of the STEREO spacecraft with respect to Earth. To this model we apply the information on CH evolution by comparing CH areas extracted in EUV data from STEREO and Earth perspective. By analyzing the evolution of CHs from different satellites we tracked the history of a CH and how changes in the CH area affect the solar wind speed arriving at 1~AU.

Compared to ACE+27 (ACE in-situ measurements shifted forward by a full solar rotation), STEREO and STEREO+CH use the most recent available information to forecast the solar wind speed. This is well reflected when comparing the ``hit and miss'' performance of the different models, revealing a clear improvement between ACE+27 and STEREO, and further on between STEREO and STEREO+CH. By taking into account the evolution of CHs, compared to ACE+27, the STEREO+CH model is able to reduce the misses by about 23\%, the false alarms by about 19\%, and increases the hit rate by about 12\%. In comparison to STEREO, STEREO+CH can reduce the misses by about 8\%, increase the hits by about 4\%, but produces a higher false alarm rate by 5\%. Especially during times of enhanced activity (as in 2012), the performance is better for STEREO+CH compared to ACE+27 and STEREO. The additional information of CH evolution is however not able to deliver more accurate solar wind speed forecasts than using the STEREO persistence model without CH information. In general, the correlation between forecast and measured speed increases when excluding CME events, hence, for operational persistence models an automatic detection of CMEs from in-situ data would be an advantage \citep[see e.g.][]{vennerstroem15}.

The STEREO+CH model produces an intensified profile for those solar wind streams that are related to expanding CHs, while suppressing those which are related to decaying CHs. This is in favor for the prediction as it increases the number of hits and reduces the number of misses. However, we find that the derived CH evolution ratio cannot be related in a simple linear way to the observed speed variations. On average, we obtain that expanding CHs more likely cause an increase in the related solar wind speed stream, while decaying CHs show less clear or no trends at all related to a decrease in the speed. Only for the year 2017 we derive distinct results showing that decaying CHs cause a decrease in the solar wind speed. On a statistical basis, for larger separation angles (and increased solar activity) we observe more decaying than expanding CHs. From this we may speculate that either the decay phase of a CH is of longer duration compared to its expansion phase or, as the solar activity picked up in 2012, more CHs decayed. An increased solar activity contributes to a more frequent and eruptive opening of magnetic field (CME events), hence, maybe a faster evolutionary phase of CHs. The behavior of CHs and solar wind evolution has a solar cycle dependence and varies with changes in the underlying as well global magnetic field \citep[e.g.][]{wang90,luhmann09,petrie13}. In that respect, also polar coronal holes influence the solar wind speed in the ecliptic plane, however, the interplay between polar and low lying coronal holes is not fully understood yet \citep[see also a recent review by][]{cranmer17}. \cite{miyake12} concluded from investigating STEREO in-situ data during 2007--2009 for predicting the Kp-index, that a simple correlation method of solar wind measurement at separated solar longitude is not enough even though the correlation is generally high. Also our study revealed that solar wind high-speed streams related to decaying CHs undergo various speed changes that cannot be covered by a simple statistical approach. As we observe a large fraction of decaying CHs, STEREO+CH is not able to deliver more accurate forecasts. In order to refine future models, the relation between solar surface structures as magnetic field and CHs, and in-situ solar wind plasma and magnetic field needs to be investigated in more detail to better understand the physics behind. 

The study shows the importance for an L5 mission covering at least EUV and magnetic imager, and in-situ plasma and magnetic field instruments, to explore in more detail the mid-term evolution of solar surface structures and their effects in the solar wind. The real-time application of the STEREO+CH persistence model and comparison to in-situ measurements can be found at $\href{http://swe.uni-graz.at}{swe.uni-graz.at}$ under \textit{Services}.


\begin{acknowledgements}
M.T. acknowledges the support by the FFG/ASAP Programme under grant no. 859729 (SWAMI). J.H.\, is supported by the CCSOM project funded by BELSPO (BRAIN-be). We thank the ACE SWEPAM instrument team and the ACE Science Center for providing the ACE data. SDO data are courtesy of the NASA/SDO and the AIA science team, STEREO data are courtesy of the NASA/STEREO and the SECCHI and PLASTIC science teams. M.A.R.\,acknowledges the Austrian Science Fund (FWF): J4160-N27. We also would like to thank the anonymous reviewers for their careful reading of our manuscript and their helpful comments. 

\end{acknowledgements}

\end{document}